# Interference of magnetointersubband and phonon-induced resistance oscillations in single GaAs quantum wells with two populated subbands

A.A.Bykov and A.V.Goran

*Institute of Semiconductor Physics, Russian Academy of Sciences, Siberian Division, Novosibirsk, Russia*

S.A.Vitkalov

*Physics Department, City College, City University of New York, New York, New York 10031, USA*

Low-temperature electron magnetotransport in single GaAs quantum wells with two populated subbands is studied at large filling factors. Magneto-inter-subband (MIS) and acoustic-phonon induced oscillations of the dissipative resistance are found to be coexisting but interfering substantially with each other. The experiments show that amplitude of the MIS-oscillations enhances significantly by phonons, indicating "constructive interference" between the phonon scattering and the intersubband electron transitions. Temperature damping of the quantum oscillations is found to be related to broadening of Landau levels caused by considerable electron-electron scattering.

The magnetotransport phenomena in high-mobility modulation-doped semiconductor structures are commonly studied with only one populated subband ($E_1$), because the electron mobility decreases with filling second subband ($E_2$) due to intersubband scattering [1]. The later also gives rise to, so-called, magneto-inter-subband oscillations (MISO) of the dissipative resistance $r_{xx}$ [2]. In electron systems with two populated subbands MISO have its maxima in magnetic fields $B$ satisfying to the relation [3-5]: $\Delta_{12} = k\hbar w_c$, where $\Delta_{12} = E_2 - E_1$ is energy separation of the subbands, $w_c = eB/m^*$ is cyclotron frequency, $m^*$ is effective electron mass and $k$ is a positive integer. Similarly to well-known Shubnikov-de Haas (SdH) oscillations MISO require quantization of the electron spectrum and are periodic in inverse magnetic fields. The electron quantum relaxation time $t_q$ determine the amplitude of the oscillations.

Another class of magnetoresistance oscillations in semiconductor structures with high electron mobility emerges at elevated temperatures, when acoustic phonon modes with Fermi momentum become to be populated [6,7]. These oscillations are called phonon-induced resistance oscillations (PIRO). PIRO are result of a modulation of the probability of the electron scattering on phonons with different momentum in quantizing magnetic fields. The dominant phonon-induced scattering between Landau levels changes the direction of the electron motion by 180 degree, providing $2k_F$ variation of the electron wave vector. The modulation, thus, induces resistance oscillations (PIRO) with the period determined by [6]: $j = 2k_F u_s/w_c$, where $u_s$ is acoustic phonon velocity and $j$ is positive integer.

In this paper we show that MISO and PIRO coexist in single GaAs quantum wells with two populated subbands. The coexistence, however, is not a simple sum of the two effects. Our data indicates significant interference between the phonon and the inter-subband scattering, which, to the best of our knowledge, has not been seen yet. The experiment shows that amplitude of the MIS-oscillations enhances significantly by phonons, indicating a nontrivial effect of the phonon scattering on the linewidth of the intersubband quantum transitions. We investigate also effects of the temperature on the amplitude of the quantum oscillations. The oscillations decay at high temperature. The temperature suppression of the oscillations is found to be consistent with the broadening of the quantum electron levels and /or with the decrease of the quantum scattering time $t_q(T)$ at high temperatures. The temperature variation of the quantum time is found to be inversely proportional to square of the temperature, indicating leading contribution of the electron-electron interaction to the decrease of the electron quantum coherence with the temperature [8,9].

In classically strong magnetic fields the dissipative resistivity $r_{xx}$ can be written as $r_{xx} \cong s_{xx}/s_{xy}^2$, where $s_{xx}$ is dissipative conductivity, $s_{xy} = e^2 n_e/m^* w_c$ is classical Hall conductivity, $n_e$ is carrier density. For the presentation of the experimental results following below we consider the total resistivity $r_{xx}$ (and conductivity $s_{xx}$) as the sum of independent components coming from different scattering mechanisms. Contributions of both the magneto-intersubband scattering and the phonon-induced scattering, which is important at high temperatures [6-9], are additively included to the total response. The resistance $r_{xx}$ can be presented as [10]:

$$r_{xx} = r_{xx}^{(0)} + r_{xx}^{(1)} + r_{xx}^{(2)}, \tag{1}$$

where $r_{xx}^{(0)}$ is quasiclassical resistance, $r_{xx}^{(1)}$ is first-order quantum contribution, $r_{xx}^{(2)}$ is second-order quantum contribution. The quasi-classical magnetoresistance reads [11]:

$$r_{xx}^{(0)} = r_0 + r_0[rn_1n_2m_1m_2(m_1 - m_2)^2 B^2]/[(n_1m_1 + n_2m_2)^2 + (rn_Tm_1m_2)^2 B^2], \quad (2)$$

where $m_1$ and $m_2$ is mobility in both subbands, $n_T = n_1 + n_2$, and $r$ is dimensionless parameters responsible for inter-subband scattering. The SdH oscillations present the first-order quantum contribution, while the MISO and PIRO are in the second-order term. The contribution of the MIS-oscillations is [5, 10]:

$$\Delta r_{MISO} = (2m^*/e^2 n_s)n_{12}\exp[-(p/w_c)(1/t_{q1} + 1/t_{q2})]\cos(2p\Delta_{12}/\hbar w_c), \quad (3)$$

where $n_{12}$ is effective inter-subband scattering rate. Positive quantum magnetoresistance is another second-order term [10]:

$$r_{QUMR} = (2m^*/e^2 n_s)[(n_1/n_s)n_{11}\exp(-2p/w_c t_{q1}) + (n_2/n_s)n_{22}\exp(-2p/w_c t_{q2})], \quad (4)$$

where $n_{11}$ and $n_{22}$ is scattering rates in subbands, $t_{q1}$ and $t_{q2}$ is quantum relaxation times in subbands.

In the case of two occupied subbands and one acoustic mode with velocity $u_s$ the phonon-induced oscillations contain four terms corresponding to both the scattering in each of two subbands ($2k_{F1}u_s = j_1 w_c$ and $2k_{F2}u_s = j_2 w_c$) and the phonon-induced intersubband scattering (($k_{F1} + k_{F2})u_s = j_3 w_c$ and $(k_{F1} - k_{F2})u_s = j_4 w_c$), where $k_{F1,2}$ is Fermi wave vector in subbands, $j_1$, $j_2$, $j_3$ and $j_4$ are positive integers. For the phonon-induced resistance oscillations $\Delta r_{PIRO}$ an analytical expression is obtained recently for a single-subband system, in which three periodic components of PIRO corresponding to the electron resonance scattering on three phonon modes in GaAs are identified [12]. The case of the two populated subbands with three acoustic modes has not been analyzed yet, but it is likely that in this case the PIRO may contain a sum of 12 periodic components.

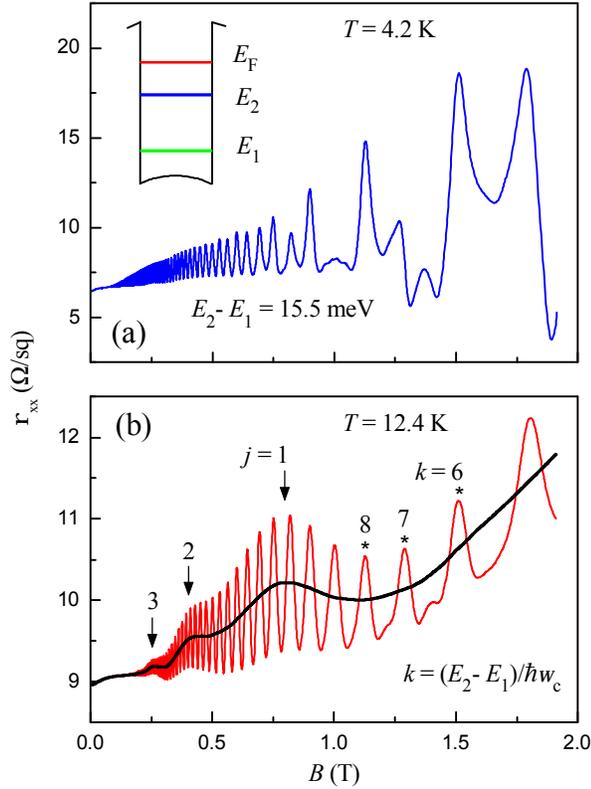

Fig. 1. (a) Experimental curve of $r_{xx}$ vs $B$ measured in GaAs quantum well with AlAs/GaAs superlattice barriers at $T = 4.2$ K. Inset presents the energy diagram of quantum well with two populated subbands with energy $E_1$ and $E_2$. (b) Experimental curve of $r_{xx}$ vs $B$ at $T = 12.4$ K (thin line), the same curve after removing high-frequency MISO (thick line). Asterisks mark the maxima of MISO numbered $k = 6$, 7 and 8. Arrows mark the maxima of PIRO numbered $j = 1$, 2 and 3.

We studied symmetrically doped single GaAs quantum wells (width $d_W = 26$ nm) with GaAs/AlAs superlattice barriers [13, 14] grown using molecular-beam epitaxy on (100) GaAs substrates. The energy diagram of quantum well with two populated subbands is shown on the inset to Fig. 1. Magnetoresistance measurements were performed on 450x50 $mm$ Hall bars fabricated using optical lithography and liquid etching in temperature range $T = 4.2 - 18.4$ K and magnetic field $B < 2$ T. Magnetoresistance $r_{xx}(B)$ and $r_{xy}(B)$ was measured on low-frequency (0.01-1 kHz) current not exceeding $10^{-6}$A. Hall concentration of electrons $n_H = 8.14 \times 10^{15}$ m$^{-2}$ was obtained from resistance measured in magnetic field $B = 0.5$ T, Hall mobility $m_H = 119$ m$^2$/Vs was calculated from $n_H$ and $r_{xx}(B = 0) = r_0$ at $T = 4.2$ K. The Fourier transform of oscillatory part of $r_{xx}(1/B)$, measured at $T = 4.2$ K (Fig. 1a), discloses two peaks corresponding to SdH oscillations (with frequencies $f_1$ and $f_2$) and one peak corresponding to MISO (with frequency $f_{MISO} \cong f_1 - f_2$) [15]. Electron concentration in subbands was calculated from frequencies $f_1$ and $f_2$: $n_1 = 2ef_1/h = 6.24 \cdot 10^{15}$ m$^{-2}$ and $n_2 = 2ef_2/h = 1.91 \cdot 10^{15}$ m$^{-2}$. The subband energy separation obtained from $f_{MISO}$ was $\Delta_{12} = 15.5$ meV.

Fig 1a presents a typical experimental curve of $r_{xx}(B)$ for studied structures at $T = 4.2$ K. The oscillatory part of $r_{xx}$ in the quantum well with two populated subbands consists of two series of SdH oscillations accompanied by MISO [2-5]. Fourier transform of $r_{xx}(1/B)$ discloses that in magnetic field range 0.01 T $< B <$ 0.5 T we observe only MISO with maxima corresponding to $k = \Delta_{12}/\hbar w_c$, while in higher magnetic fields MISO coexist with SdH oscillations. At temperature $T = 12.4$ K SdH oscillations are completely damped due to temperature broadening of the electron distribution. The amplitude of MISO decreases with increasing temperature, but MISO can still be clearly observed in wide range of magnetic fields B > 0.2 T. In addition to MISO in Fig. 1b there is another oscillating component presented by the thick line. The asterisks mark the maxima of MISO with $k = 6, 7$ and 8, while the arrows mark three maxima of the new oscillating component.

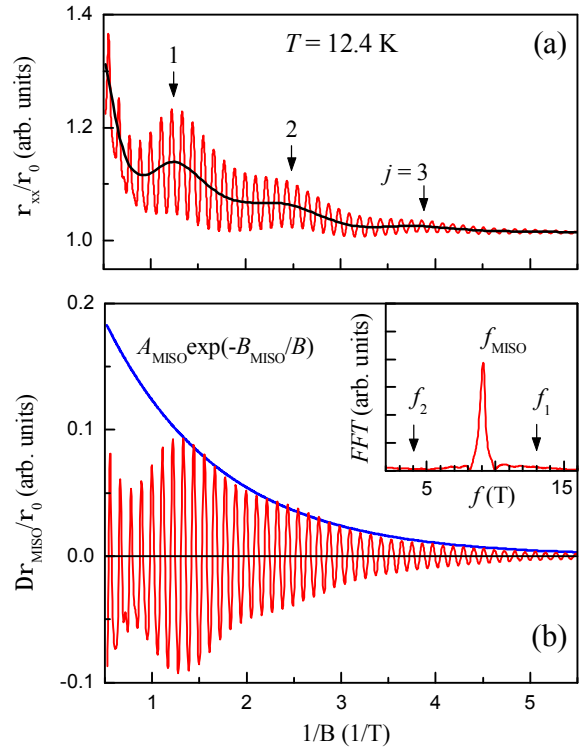

Fig. 2. (a) Experimental curve of $r_{xx}/r_0$ vs $1/B$ measured in GaAs quantum well with two populated subbands at $T = 12.4$ K (thin line) and the same curve after removing high-frequency MISO (thick line). Arrows mark the maxima of PIRO numbered $j = 1, 2$ and 3. (b) Experimental curve of $Dr_{MISO}/r_0$ vs $1/B$ (thick line) and calculated curve $A_{MISO} \exp(-B_{MISO}/B)$ with $A_{MISO} = 0.28$ and $B_{MISO} = 0.82$ Тл (thick line). Inset presents the Fourier transform of $r_{xx}/r_0$ vs $1/B$. Arrows mark the frequencies of SdH oscillations in subbands.

The Fig 2a shows the relative magnetoresistance $r_{xx}/r_0(1/B)$ versus inverse magnetic fields at $T = 12.4$ K. The inset to Fig 2b presents the Fourier transform of the oscillating part of $r_{xx}(1/B)$ measured at $T = 12.4$ K. In the inset the peak corresponds to MIS-oscillations. There are no peaks at the frequencies of

SdH oscillations ($f_1$ = 12.9 T$^{-1}$ and $f_2$ = 3.95 T$^{-1}$ pointed by arrows on the inset). At T=12.4 K the contribution of SdH oscillations (the first-order term in Eq. (1)) to the magnetoresistance is negligibly small. Fig 2b shows an extracted MISO content, which is difference between the high frequency oscillations (the thin line in Fig 2a) and the low frequency background (the thick line). The amplitude of MISO rises with increasing magnetic field up to $B$ = 0.8 T. It can be well approximated by an expression $A_{MISO}\exp(-B_{MISO}/B)$ where $A_{MISO}$ and $B_{MISO}$ are fitting parameters. The approximation follows from the theory [5, 10].

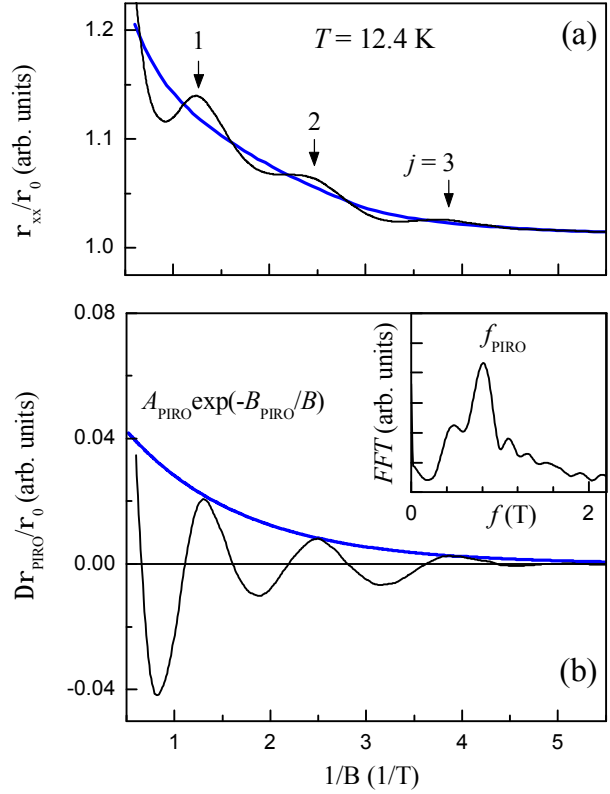

Fig. 3. (a) Experimental curve of $r_{xx}/r_0$ vs $1/B$ after removing MISO (thin line) and its monotonous component (thick line) at $T$ = 12.4 K. Arrows mark the maxima of PIRO numbered $j$ =1, 2 and 3. (b) Experimental curve of $Dr_{PIRO}/r_0$ vs $1/B$ (thin line) and calculated curve $A_{PIRO} \exp(-B_{PIRO}/B)$ with $A_{PIRO}$ = 0.064 and $B_{PIRO}$ = 0.82 Тл (thick line). Inset presents the Fourier transform of $Dr_{PIRO}/r_0$ vs $1/B$.

At higher magnetic field $B$ > 0.8 T ( $1/B$<1.25 T$^{-1}$) the amplitude of MIS-oscillations decreases significantly and deviates from the theory. The amplitude modulation correlates well with the behavior of the low frequency background: decrease of the MISO amplitude occurs at minima of the low frequency background (compare Fig. 2a and 2b). As shown below the oscillating background corresponds to the phonon-induced oscillations of the resistivity (PIRO). Thus the modulation of the MISO amplitude indicates a significant interference between these two phenomena. At first glance, since the quantum oscillations are sensitive to the quantum scattering time $t_q$, the strong decrease of the MISO amplitude at $1/B$<1.2 T$^{-1}$ could be a result of the decrease of the quantum scattering time of electrons, induced by intense electron-phonon scattering. However the decrease of the MISO amplitude occurs at the minimum of the PIR-oscillation (see Fig.2a), where, apparently, the rate of the electron-phonon scattering is at a minimum and, therefore, the quantum scattering time $t_q$ should be longer. Thus, instead expected "destructive interference" between two processes, the data indicates a "constructive interference" between MISO and PIRO. Namely the enhancement (decrease) of the phonon-induced electron transitions increases (decreases) the amplitude of the inter-subband magnetooscillations. This indicates a phonon-induced "effective" narrowing of the intersubband resonance in the electron scattering.

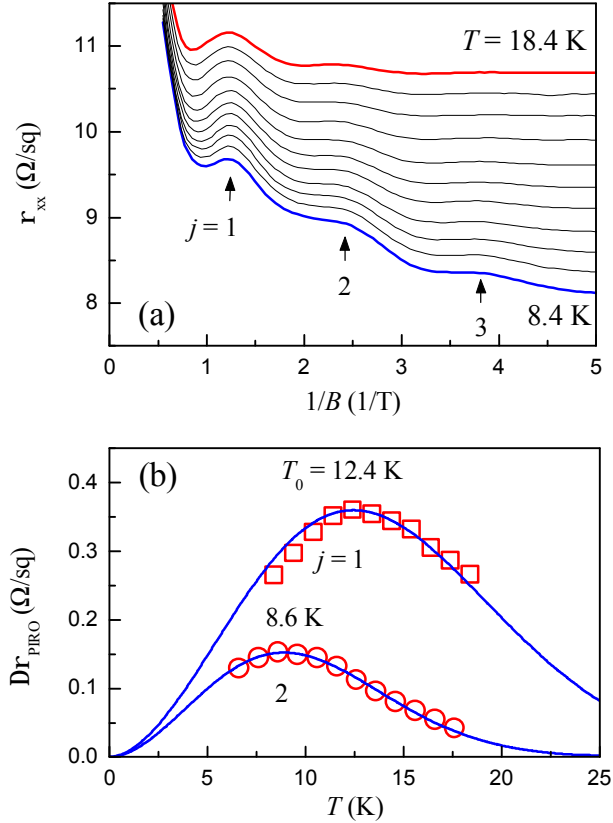

Fig. 4. (a) Experimental curves of $r_{xx}$ vs $1/B$ after removing MISO at different temperatures in range $8.4 < T < 18.4$ K. Arrows mark the maxima of PIRO numbered $j$ =1, 2 and 3. (b) Experimental (squares and circles) and calculated according to (5) (solid lines) curves of $Dr_{PIRO}$ vs $T$ for $j$ =1 and 2.

A dynamic reduction of resonance linewidth is known phenomenon in the area of spin resonance [16]. In many cases the linewidth of a spin resonance decreases with an increase of the temperature. The linewidth reduction is result of an effective averaging of stochastic dipole interaction between spins, when the spins diffuse spatially, exploring more of the phase space. We suggest that a similar physics could be responsible for the apparent decrease of the Landau level width with the enhancement of the electron-phonon scattering. At high filling factors the level width is determined by an averaging out of the random smooth potential seen by an electron moving along a cyclotron orbit [17]. The electron-phonon scattering enhances the electron diffusion, providing a better averaging out of the impurity potential and, thus, reducing the width of Landau levels.

Below we consider the behavior of the low frequency oscillations of the resitivity. Fig. 3a presents relative magnetoresistance $r_{xx}/r_0(1/B)$ after removing high-frequency MISO (thin line). It can be separated to monotonous component (thick link on Fig 3a) and low-frequency oscillating component (thin line on Fig 3a) that only appears at high temperature ($T = 12.4$ K) and comes from magneto-phonon resonance in high-mobility heterostructures at large filling factor [6]. The Fourier transform of low-frequency oscillating component is shown on the inset to Fig. 3b. Assuming that the main peak on Fourier transform (with frequency $f_{PIRO}$) corresponds to the resonance scattering of electrons on acoustic phonons in first subband we obtained the value of $u_s = 5.2$ km/s, which is consistent with $u_s$ for one of the bulk acoustic modes in (100) GaAs layers [12]. It's worth mentioning that the $j(1/B)$ is not strictly linear. One of the reasons for this is that all three bulk acoustic modes participate in resonance scattering. Another reason is that in system with two populated subbands the contributions to $r_{xx}$ from both subbands and from inter-subband scattering should be taken into account.

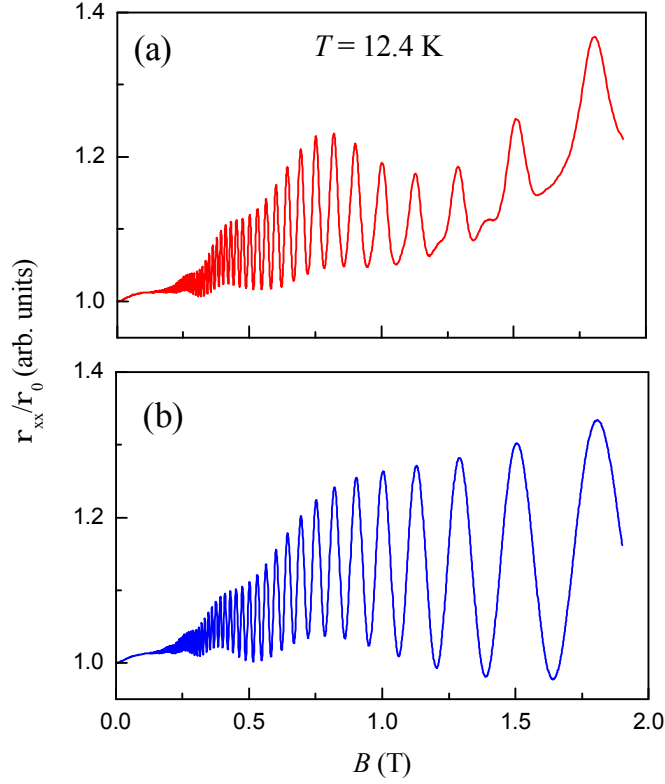

Fig. 5. (a) Experimental curves of $r_{xx}/r_0$ vs $B$ measured in GaAs quantum well with AlAs/GaAs superlattice barriers at $T$ = 12.4 K. (b) Calculated curve of $r_{xx}/r_0$ vs $B$ in quantum well with two populated subbands.

Fig. 4a presents magnetoresistance $r_{xx}(1/B)$ after removing high-frequency MISO for temperatures in range 8.4 - 18.4 K. The first oscillation ($j$ = 1) is clearly seen for all temperatures, while the second and third oscillations ($j$ = 2, 3) disappear with increasing temperature. The $Dr_{PIRO}(T)$ is presented on Fig. 4b and has maxima at $T_0$ = 12.4 K for $j$ = 1 and $T_0$ = 8.6 K for $j$ = 2, which is consistent with the results obtained in [8, 9]. The $Dr_{PIRO}(T)$ for $j$ = 3 cannot be calculated with a decent precision. The solid lines on Fig. 4b were calculated as [8]:

$$Dr_{PIRO}(T) \propto t_{ph}^{-1}(T)\exp[-2p/w_c t_q^{ee}(T)], \tag{5}$$

where $t_{ph}$ is relaxation time for scattering on acoustic phonons, $t_q^{ee}$ is quantum lifetime for $e$-$e$ scattering. The term $t_{ph}^{-1}(T)$ is responsible for rise of PIRO with increasing temperature, $\exp(-2p/w_c t_q^{ee})$ is responsible for its temperature damping. Assuming that $1/t_{ph}(T) \propto T^\alpha$ [18-20] and $1/t_q^{ee}(T) = lT^2/E_F$ [21, 22] the $Dr_{PIRO}(T)$ can be expressed using two fitting parameters $a$ and $l$ [8]. In our calculation we used the value of $a$ = 1.8 [8] and $E_F = E_{F1}$. We obtained a good agreement between experimental and theoretical curves for $l$ = 3.8.

Fig. 5 presents the experimental and calculated dependences of $r_{xx}/r_0(B)$ at $T$ = 12.4 K. We calculated $r_{xx}/r_0(B)$ by summing the contributions of all scattering mechanisms. The quasiclassical magnetoresistance $r_{QCMR}/r_0$ was calculated using (2). The good agreement with experimental data was obtained for the following parameters: $m_1$ = 88.4 m$^2$/Vs, $m_2$ = 76.6 m$^2$/Vs, $r$ = 0.25. The value of $r$ indicates the importance of inter-subband scattering in studied system. The values of $m_1$ and $m_2$ were close to $m_0 = 1/er_0 n_s$ with an accuracy of 10%, which allows us to consider the scattering rates in the subbands to be almost the same. After such simplification the magnetoresistance in (4) can be written as: $r_{QUMR}/r_0 = A_{QUMR} \exp(-B_{QUMR}/B)$ where $A_{QUMR}$ and $B_{QUMR}$ are fitting parameters. The same way we included oscillating components coming from the inter-subband scattering and oscillating components from scattering on acoustic phonons: $\Delta r_{MISO}/r_0 = A_{MISO}\exp(-B_{MISO}/B)\cos(2p\Delta_{12}/\hbar w_c)$, $\Delta r_{PIRO}/r_0 = A_{PIRO}\exp(-B_{PIRO}/B)\cos(4pk_{F1}u_s/w_c)$. The fitting parameters $A_{MISO}$, $B_{MISO}$, $A_{PIRO}$, $B_{PIRO}$ were calculated with the assumption that the scattering rates are the same in both subbands, which is valid at $T$ = 12.4 K: $B_{QUMR}$ =

$B_{MISO} = B_{PIRO} = 0.82$ T, $A_{QUMR} = A_{MISO} = 0.28$, $A_{PIRO} = 0.064$. The calculated resistance $r_{xx}/r_0(B)$ matches the experimental curve in low magnetic field $B < 0.7$ T. The correspondence breaks at higher magnetic field, since the calculations do not account the substantial interference between the MISO and PIRO. Additional analysis is required to explain and describe quantitatively the observed effect.

In conclusion, we have found that the magneto-intersubband oscillations of dissipative resistance in single GaAs quantum wells with two populated subbands coexist with the phonon-induced oscillations. These two kinds of the quantum oscillations interfere constructively. The experiment demonstrates that the enhancement of the electron-phonon scattering increases the amplitude of MISO, indicating decrease of the linewidth of the intersubband resonance. We suggest that the effect is similar to the diffusive narrowing of spin resonance. The temperature suppression of the oscillations is found to be consistent with the broadening of the quantum electron levels and /or with the decrease of the quantum scattering time at high temperatures. The temperature variation of the quantum time is found to be inversely proportional to square of the temperature, indicating leading contribution of the electron-electron interaction to the decrease of the electron quantum coherence with the temperature.

The authors thank L. I. Magarill for useful discussion. The work was supported by RBFR project #08-02-01051 and #10-02-00285.